\pdfoutput=1
\documentclass{JINST}

\title{A simple and robust method to study after-pulses in Silicon Photomultipliers}

\author{M. Caccia\thanks{Corresponding author.}, R. Santoro, G. A. Stanizzi \\
\llap{}Dipartimento di Scienza e Alta Tecnologia, Universita' degli Studi dell'Insubria,\\
  Via Valleggio 11, 22100, Como, Italy.\\
  E-mail: \email{massimo.caccia@uninsubria.it}}

\abstract{The after-pulsing probability in Silicon Photomultipliers and its time constant are obtained measuring the mean number of photo-electrons in a variable time window following a light pulse. The method, experimentally simple and statistically robust due to the use of the Central Limit Theorem, has been applied to an HAMAMATSU MPPC S10362-11-100C.}


\begin{document}

\section{Introduction}

Silicon Photomultipliers (SiPM) are state-of-the-art  detectors of light  consisting of a matrix of P-N junctions with a common output, with a density of cells up to $\approx 10^{4}/mm^{2}$. Each diode is operated in a limited Geiger-Muller regime in order to achieve gains at the level of $\approx 10^{6}$ and to guarantee an extremely high homogeneity in the cell-to-cell response. Subject to the high electric field in the depletion zone, initial charge carriers generated by an absorbed photon or by thermal effects trigger an exponential charge multiplication by impact ionization. When the current spike across the quenching resistor induces a drop in the  voltage across the junction, the avalanche is stopped. SiPM can be seen as a collection of binary cells, providing altogether an information about the intensity of the incoming light by counting the number of fired cells \cite{Dolgoshein} -\cite{Piemonte}.\\

SiPM feature an unprecedented photon number resolving capability and offer relevant advantages due to the low operating voltage, the immunity to magnetic field, ruggedness and the design flexibility due to the Silicon Technology. However, they also suffer from drawbacks related to a significant temperature dependence of the gain and a high rate of spurious hits. The latter is due to thermally generated carriers (Dark Counts, DC), Optical Cross-Talk (OCT) and after-pulsing. The OCT is linked to photons generated during a primary avalanche, triggering simultaneous secondaries \cite{Renker},\cite{Lacaita}. The OCT is affected by the sensor design \cite{Renker},\cite{Nagano} -\cite{Pagano} and strongly depends on the bias voltage. After-pulses are associated to the late release of a charge carrier that has been produced in the original avalanche and trapped by an impurity \cite{Renker}. After-pulsing is essentially dependent on the sensor technology \cite{Renker},\cite{Nagano},\cite{Pagano}. 

Dark Counts, Optical Cross-Talk and after-pulsing occur stochastically and introduce fluctuations in the multiplication process that contribute to deteriorate the resolution in both photon counting and spectrometry. Moreover, after-pulsing may be critical  in photon correlation experiments \cite{Niggemann} -\cite{Toyama}.\\    

The after-pulsing effect has been investigated by various authors, relying on the time correlation of neighbouring pulses \cite{Eckert} -\cite{Piemonte2}. A simple and statistically robust method is proposed here,  based on the sensor current integration and the use of the Central Limit Theorem to estimate the mean number of pulses in a variable time window.

\section{Materials and methods}

\begin{figure}
\centering
 \includegraphics[width=.6\textwidth]{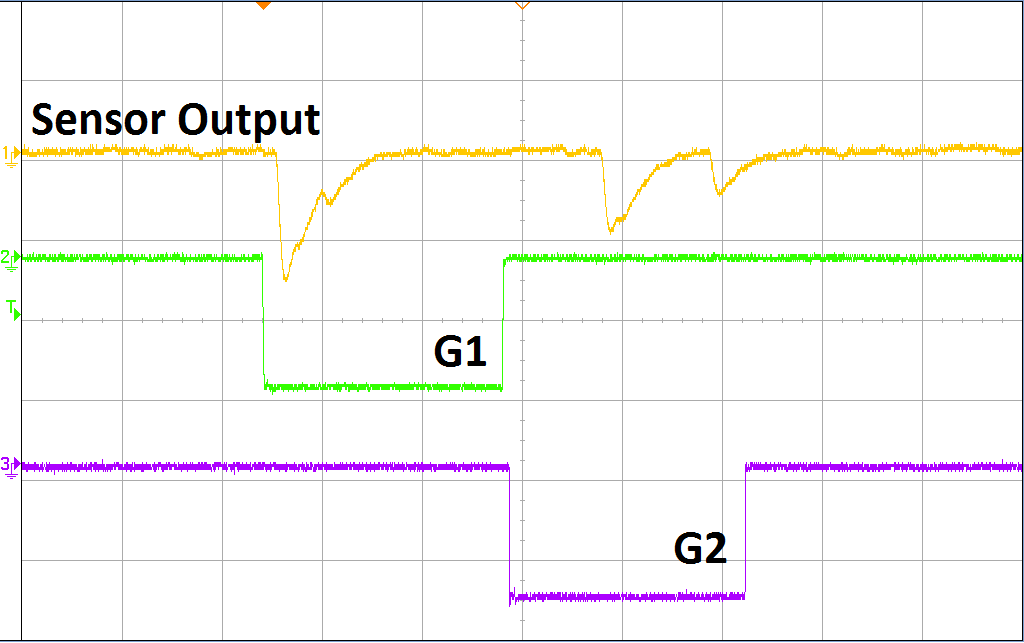}
 \caption{The figure shows the response of the SIPM to a light burst, synchronised with the leading edge of the gate $G_{1}$. Pulses in the $G_{2}$ gate may be due to both random Dark Counts and after-pulses. The exemplary event shown here features as well an after-pulse occurring during the recovery time of the sensor. The rate of Dark Counts is measured switching off the LED.}
   \label{fig:schema}
\end{figure}

After-pulses can be seen as an excess of fired cells in a time window following the signal due to a light pulse (Figure \ref{fig:schema}), where the excess is calculated with respect to Dark Counts occasionally appearing.  A statistical analysis of the excess, varying the gate length, is expected to lead to a measurement of the after-pulsing probability and of its time constant.

\subsection{Experimental setup}

\begin{figure}
\centering
 \includegraphics[width=.6\textwidth]{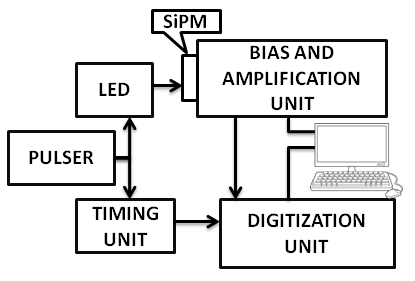}
 \caption{Block diagram of the experimental setup for the after-pulsing characterization.}
  \label{fig:moduli}
\end{figure}

The block diagram of the experimental set-up is shown in Figure \ref{fig:moduli}. A master clock is synchronising the light pulser illuminating the SiPM and the data acquisition, integrating the signal pulses in a variable duration gate $G_{2}$ delayed with respect to the light pulse. Results reported here were obtained with the following specific system:

\begin{itemize}
\item An ultra-fast LED source (SP5601 - CAEN), emitting $\approx~5~ns$ long light pulses at $405~nm$, with intensities in the 1-2000 photon range.
\item The SP5600 - CAEN  power supply and amplification unit, housing an Hamamatsu  MPPC S10362-11-100C.
\item A charge digitisation unit, either based on the CAEN - DT5720A waveform digitizer or on the CAEN - V792N QDC. 
\item A Dual-Timer  (N93B - CAEN) to generate the integration gate.
\end{itemize}

\subsection{Experimental procedure}

\begin{figure}
\centering
 \includegraphics[width=.6\textwidth]{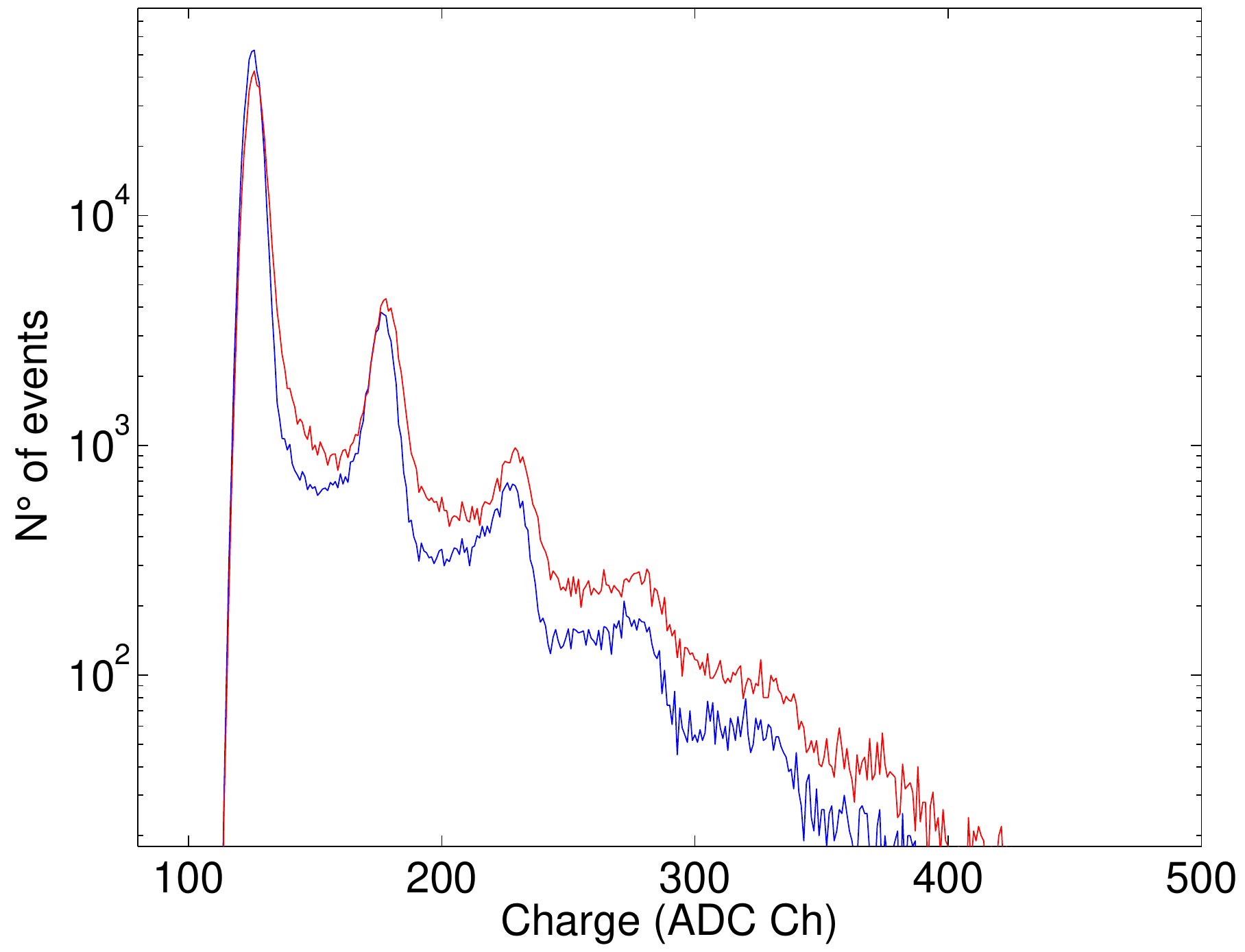}
 \caption{Spectra of the charge collected in $G_{2}$, recorded without illuminating the sensor (blue) and after a light burst (red) pulsed $500~ns$ before the gate opening. Spectra were normalised to the same number of events.}
  \label{fig:picchi}
\end{figure}

The experimental procedure  starts by defining two gates, synchronized to the light pulse (see Figure \ref{fig:schema}). The first gate $G_1$ encompasses the time development of the signal due to a light pulse. The de-trapping of charge carriers within  $G_1$ originates avalanches piled-up with the signal from the light pulse and the probability for this to occur is inferred by studying  the excess of pulses in a second variable gate $G_2$, following $G_1$. 
The estimation of the mean number of fired cells in  $G_2$ is obtained by the analysis of the spectrum of the released charge, provided integrating the sensor output current. Exemplary spectra for $G_2=400~ns$ are shown in Figure \ref{fig:picchi}, with the LED switched ON and OFF. The peak positions identify the value of the digitized charge for the different number of avalanches while the peak areas measure the corresponding probability.\\ 

\begin{figure}
\centering
 \includegraphics[width=.6\textwidth]{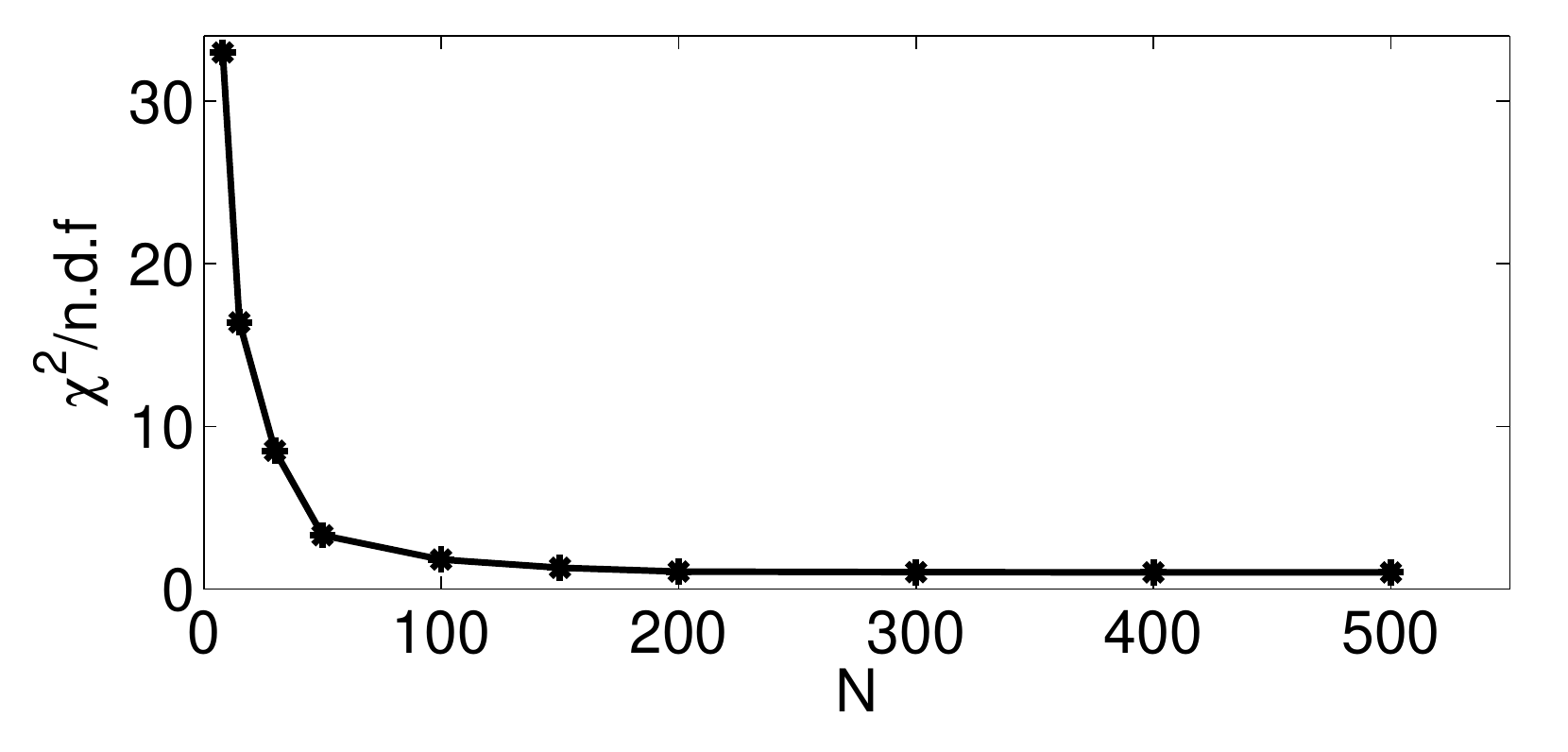}
 \caption{The $\overline{Q}_N$ distribution was tested against the hypothesis of being Gaussian. The plot shows the $\chi^2/n.d.f.$ vs. N of the fit.}
  \label{fig:chi2}
\end{figure}

\begin{figure}
\centering
 \includegraphics[width=0.6\textwidth]{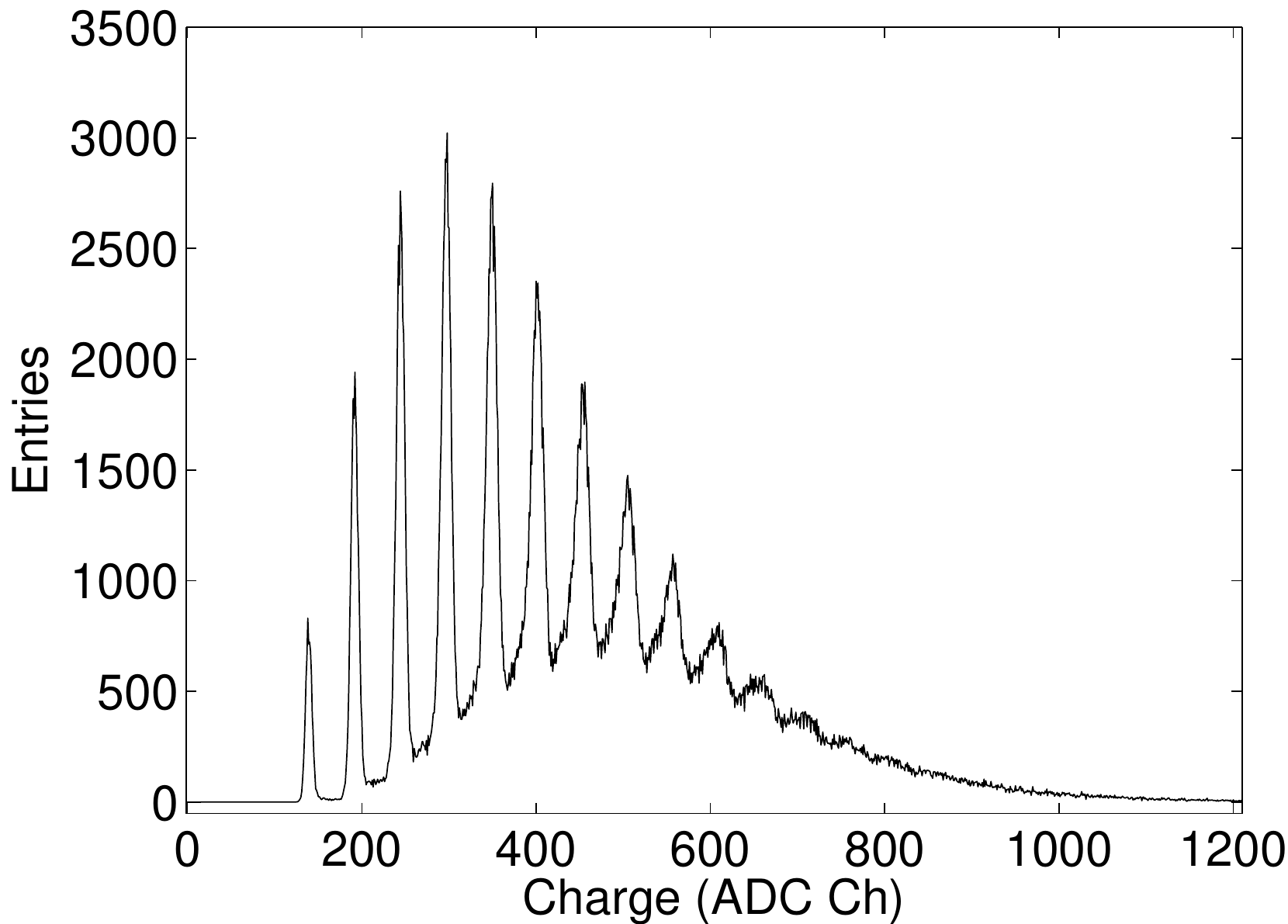}
 \caption{The distribution of the detected photons used to calibrate the charge in photo-electrons.}
  \label{fig:mgf}
\end{figure}

The spectra clearly show evidence of two statistics characterized by a different mean value. In order to obtain a quantitative information, the data were analysed as follows:

\begin{itemize}
\item The average charge $\overline{Q}_N$ by a sequence of N events was retained as the main observable. According to the Central Limit Theorem, $\overline{Q}_N$ is expected to be Gaussian distributed, with the advantage of an easy and robust way to estimate its value and to measure its uncertainty. The value of N was determined studying the evolution of the $\overline{Q}_N$ distribution vs. N and fixing its value at 200, when it was shown to be asymptotically gaussian (Figure \ref{fig:chi2}).

\item In order to turn $\overline{Q}_N$  into a number of fired cells, a multi-photon  spectrum was recorded illuminating the sensor within $G_1$. The reference spectrum is shown in Figure~\ref{fig:mgf}. This can be fit with a sum of gaussians \cite{Caccia} to find the average peak-to-peak distance $\overline{\Delta}_{pp}$ which provides the conversion factor from ADC channels to number of cells.
\end{itemize}

\begin{figure}
\centering
 \includegraphics[width=.6\textwidth]{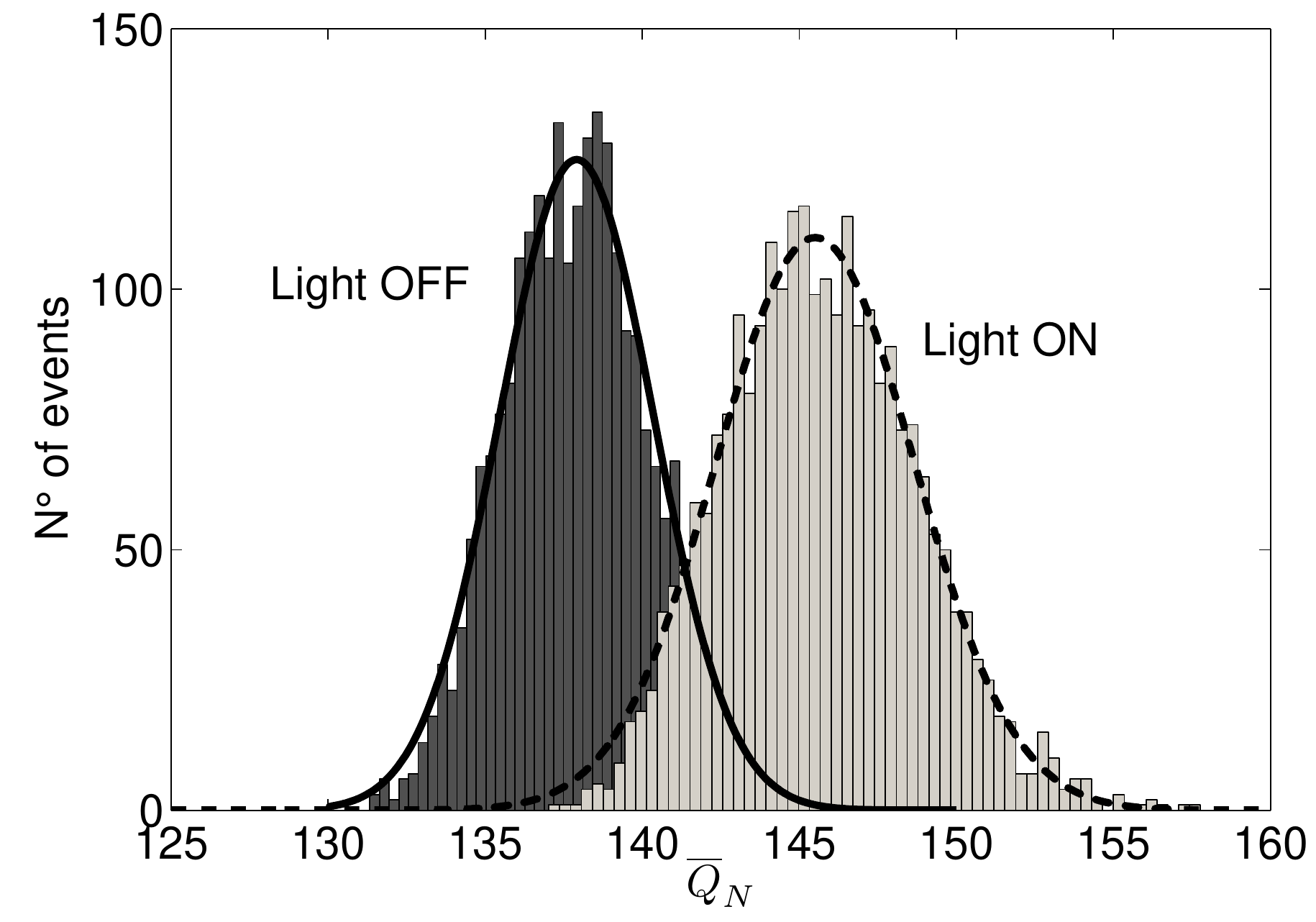}
 \caption{Exemplary distribution of $\overline{Q}_{N}$ for $G_{2}=400 ns$, having the LED switched ON and OFF.}
  \label{fig:tlc}
\end{figure}

The spectra of $\overline{Q}_N$ are shown in Figure \ref{fig:tlc}, where the shift for the illuminated sensor is very clear. Eventually, the quantity 

\begin{equation}
	\Delta_{QQ}(G_{2}) = \frac{\langle \overline{Q}_{N}(light~ON,G_{2})\rangle-\langle \overline{Q}_{N}(light~OFF,G_{2})\rangle}{\overline{\Delta}_{pp} }
	\label{eqn:pe}
\end{equation}

measures the excess of avalanches due to after pulses associated to the light burst with respect to Dark Counts.\\ 

The procedure is iterated increasing $G_2$ till when $\Delta_{QQ} (G_2)$ achieves a constant value, indicating that the after-pulsing phenomenon is exhausted.

In order to cope with possible temperature changes during the experiment, for every value of $G_2$, a multi-photon spectrum is recorded to have an actual value of $\overline{\Delta}_{pp}$.

\section{Results}

\begin{figure}
\centering
 \includegraphics[width=0.6\textwidth]{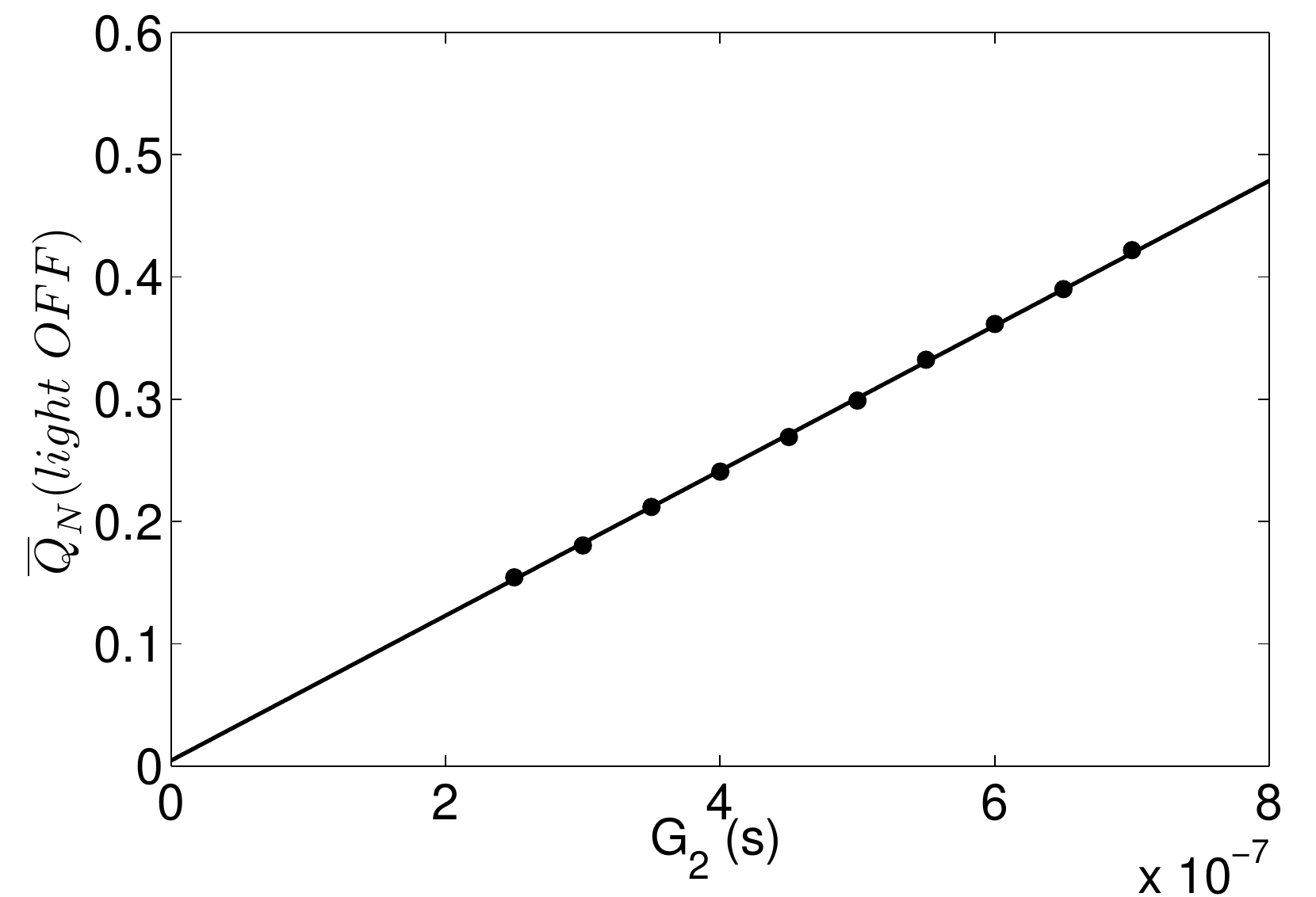}
 \caption{$\overline{Q}_{N}(light~OFF)/\overline{\Delta}_{pp}$ as a function of $G_2$ (s).}
  \label{fig:time_nl1}
\end{figure}

The procedure has been initially qualified measuring the Dark Count Rate (DCR) by the calculation of $\langle \overline{Q}_N \rangle$. The trend of $\langle \overline{Q}_{N}(light~OFF) \rangle/\overline{\Delta}_{pp}$ vs. time is shown in Figure \ref{fig:time_nl1}, where the straight line fit corresponds to a slope   $m=593~\pm~5~kHz$. 
The  average number of photo-electrons is actually affected by the OCT and $\langle \overline{Q}_N \rangle/\overline{\Delta}_{pp} = n_{p.e.}\times (1+\epsilon)$, where  $\epsilon=(22~\pm~1)\%$ is the measured Cross-Talk probability and $n_{p.e.}$ is the number of primary avalanches.  As a consequence, the DCR may be calculated  as $\frac{m}{1+\epsilon}=486~\pm~8~kHz$, in fair agreement with the value of $480~\pm~4~kHz$ by a direct count of the pulses above the 0.5 photo-electron threshold.\\

Concerning after-pulses, once the de-trapping of the charge carriers  is assumed to have an exponential time dependence, the 
probability density function may be written as:

\begin{equation}
	y(t) = \frac{P}{\tau}e^{-\frac{t}{\tau}},
	\label{eqn:pe3}
\end{equation}

where \textbf{P} is the probability for a single avalanche to originate an after-pulse and  \textbf{$\tau$} is the characteristic  time constant of the phenomenon.
The  quantity $\Delta_{QQ} (G_2)$ corresponds to the cumulative distribution function in the integration gate $G_2$, since:

\begin{equation}
	\Delta_{QQ} (G_2) = \int_{G1}^{G1+G2} \frac{N\times P}{\tau}e^{-\frac{t}{\tau}} dt = a(1-e^{-\frac{G_2}{\tau}}),
	\label{eqn:pe4}
\end{equation}
where $N$ is the mean number of photo-electrons generated by the light burst and $a=N\times Pe^{-\frac{G_1}{\tau}}$ is the asymptotic value of $\Delta_{QQ} (G_2)$.

\begin{figure}
\centering
 \includegraphics[width=0.6\textwidth]{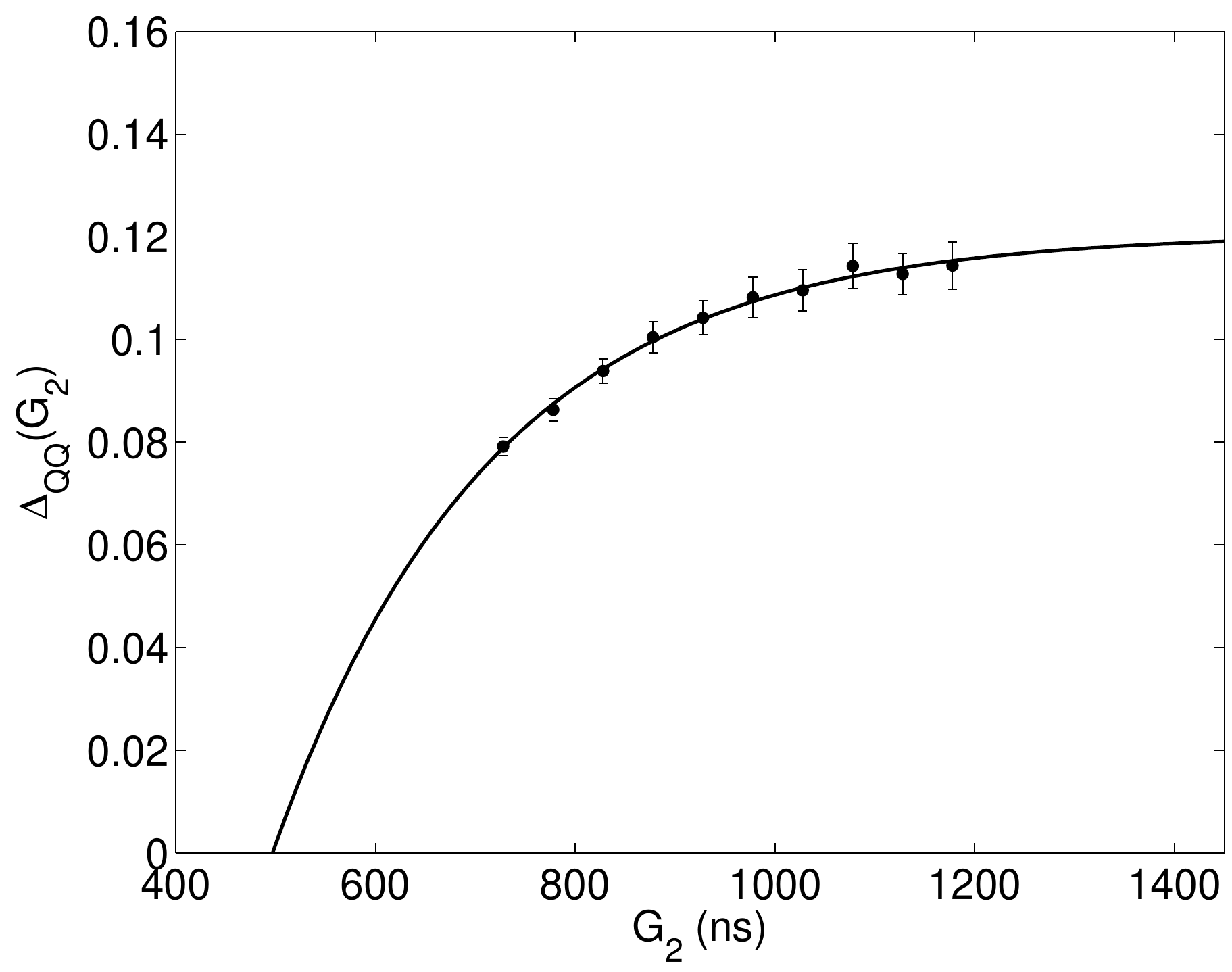}
 \caption{$\Delta_{QQ} (G_2)$ vs. $G_2$. Data are fitted by  equation 3.2, with a $\chi^2/n.d.f.=0.94$ }
  \label{fig:time_ap}
\end{figure}

The data resulting by a $G_2$ scan are shown in Figure \ref{fig:time_ap}. A fit to equation (\ref{eqn:pe4}) yields a value of $\tau=217.5~\pm~5.9~ns$ and $a~=~0.1206~\pm~0.0010$ photo electrons (p.e.). Since N was measured to be $6.1~\pm~0.1~p.e.$, this is resulting in an after-pulsing probability of $(19.44~\pm~1.38)\%$ in fair agreement with data found in the Hamamatsu datasheet \cite{hama} and in reference \cite{Niggemann},\cite{Eckert}, reporting values between 17\% and 22\%. 

\section{Conclusions}

A method for the characterisation of the after-pulses based on the analysis of the charge distribution in a variable time window has been proposed and qualified. Its advantages are the simplicity and the robustness. Its main limitation is the intrinsic impossibility to probe the after-pulsing components characterised by a short time constant.

\end{document}